\documentclass[letterpaper,twocolumn,10pt]{article}
\usepackage{usenix}

\usepackage{adjustbox}
\usepackage{algorithm}
\usepackage{algpseudocode}
\usepackage{amsmath}
\usepackage{amsfonts}
\usepackage{booktabs}
\usepackage{float}
\usepackage{graphicx}
\usepackage{mathtools}
\usepackage{pifont}
\usepackage{titlesec}

\newcommand{\Octopus}{\textsc{Swarm}}

\begin{document}

\title{\Large \bf \Octopus{}: Co-Activation Aware KVCache Offloading Across Multiple SSDs}

\author{
{\rm Tuowei Wang\thanks{These authors contributed equally.}} \\
Tsinghua University
\and
{\rm Liyun Chu\footnotemark[1]} \\
Tsinghua University
\and
{\rm Ruwen Fan} \\
Tsinghua University
\and
{\rm Ju Ren\thanks{Corresponding author (renju@tsinghua.edu.cn)}} \\
Tsinghua University
}

\maketitle

\begin{abstract}
The key-value (KV) cache has become the dominant contributor to memory consumption in large language model (LLM) inference. Although offloading KVCache from GPU high-bandwidth memory (HBM) to CPU DRAM alleviates device memory pressure, DRAM remains capacity-limited and costly for large, persistent workloads. Solid-state drives (SSDs) provide a cost-effective alternative, but naive SSD-based paging is fundamentally bandwidth-bound due to limited PCIe throughput and per-device bandwidth constraints.

In this paper, we observe that KVCache activations in real-world workloads exhibit strong and stable correlations. We term this phenomenon \textit{KVCache Co-Activation}, where accessing a KV entry is often accompanied by a stable and recurring set of other KV entries. Leveraging this property, we present \textbf{\Octopus{}}, an SSD-based KVCache offloading system that converts bandwidth-bound single-device access into parallel I/O across multiple SSDs. Specifically, \Octopus{} clusters co-activated KV entries offline and distributes the resulting clusters across SSDs using graph-based placement with selective replication to maximize parallel I/O bandwidth. At runtime, \Octopus{} performs load-balanced cluster retrieval and dynamically adapts clustering and caching decisions to sustain high bandwidth utilization under evolving access patterns. Evaluations show that \Octopus{} reduces I/O time by $2.41\times$ and improves effective bandwidth utilization by $2.72\times$.
\end{abstract}

\section{Introduction}
The key-value (KV) cache is a central component of modern large language model (LLM) inference. By storing previously computed keys and values, it eliminates redundant computation and enables efficient autoregressive decoding. However, KVCache grows monotonically with sequence length and model depth, with a lifetime that extends far beyond a single decoding step. As context lengths expand~\cite{loogle,swe-bench,next-qa} and models deepen~\cite{gemini3-pro,gpt-5,kimi-k2,deepseek-v3}, KVCache has emerged as the \textit{dominant} contributor to inference-time memory footprint.

Beyond its per-request footprint, KVCache increasingly persists across requests and sessions. For example, long documents or videos are commonly queried repeatedly with different questions~\cite{rag,video-llm}. Similarly, agentic-style interactions often reuse tool instructions~\cite{tool-llm,tool-use} and contexts~\cite{react,tree-of-thoughts} from previous steps. Real-world production deployments~\cite{prompt-caching-claude,prompt-caching-openai} have reported the widespread presence of long and shared KVCache. Consequently, KVCache has evolved from a transient artifact into a \textit{long-lived} data structure, making it a natural and compelling candidate for persistent storage. As a result, designing a scalable and efficient storage system tailored for KVCache management has become increasingly critical.

Given the limited capacity and high cost of GPU-side high-bandwidth memory (HBM), KVCache is typically offloaded to CPU-side dynamic random-access memory (DRAM)~\cite{flexgen,prompt-cache,rag-cache}. To reduce the I/O overhead between DRAM and HBM, recent work~\cite{h2o,quest,infllm,pqcache} exploits the inherent sparsity of attention by \textit{selectively activating and transferring only the most relevant KVs} for computation. Nevertheless, DRAM capacity remains limited and expensive~\cite{dram-price-1,dram-price-2}, rendering it insufficient for hosting large, persistent KVCache at scale.

In this context, \textbf{solid-state drives} (SSDs) offer an attractive alternative for KVCache offloading. Compared to DRAM, SSDs provide orders-of-magnitude higher capacity at lower cost per bit, making them well-suited for hosting large, persistent KVCache states. This cost-capacity advantage has motivated growing interest in SSD-based KVCache storage as a path toward scalable inference~\cite{impress,attention-store,mooncake,adapt-cache}. Unfortunately, naively paging KVCache between SSDs and DRAM proves ineffective~\cite{instinfer,io-aware-kvcache}. Unlike DRAM, SSDs are accessed through relatively narrow PCIe interfaces and provide limited per-device bandwidth~\cite{impress}. As a result, KVCache accesses served from SSDs become \textbf{bandwidth-bound}, stalling attention computation and increasing inference latency.

In this paper, we present a key observation that KVCache in LLMs exhibits strong correlations in their activation patterns, a phenomenon we term \textbf{\textit{KVCache Co-Activation}}. This property, prevalent in LLM workloads yet largely underexplored, can be strategically leveraged to alleviate SSD bandwidth bottlenecks during KVCache offloading. Specifically, when processing real-world datasets, the activation of an individual KV pair is consistently accompanied by a stable and recurring set of other KV pairs. These co-activated KVs can be fetched simultaneously, naturally exposing I/O parallelism that can be exploited to improve effective SSD bandwidth.

\begin{figure}[t]
    \centering
    \includegraphics[width=1.0\linewidth]{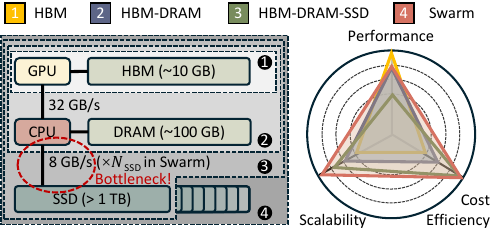}
    \caption{Memory hierarchy for KVCache offloading. \Octopus{} exploits KVCache co-activation to aggregate multi-SSD bandwidth for better performance–scalability trade-offs.}
    \label{fig:core-result}
\end{figure}

Motivated by this observation, we propose \textbf{\Octopus{}}, a novel SSD-based solution for KVCache offloading. Rather than relying on a single SSD and being constrained by per-device bandwidth, \Octopus{} offloads KVCache across \textit{multiple SSDs} and strategically aggregates bandwidth across all devices. A key idea in \Octopus{} is to align KVCache co-activation patterns with multi-SSD I/O parallelism: \textit{KV pairs that are frequently co-activated are identified, distributed, retrieved, and updated across SSDs to maximize parallel reads}. As shown in Figure~\ref{fig:core-result}, \Octopus{} transforms SSD-based KVCache offloading from a bandwidth-bound, single-device operation into a highly parallel, multi-device data path, enabling efficient and scalable KVCache management.

\Octopus{} employs a two-phase design that incorporates hierarchical optimizations performed both offline and online:

\textbf{In the offline phase}, \Octopus{} captures KVCache co-activation patterns and organizes KV entries into correlation-aware clusters, while building a lightweight DRAM-resident index using cluster medoids as representatives. Guided by this clustered layout, it partitions KVCache across the DRAM-SSD hierarchy by retaining medoids, local-window entries, and hot clusters in DRAM, and distributing the remaining clustered entries across SSDs to enable parallel retrieval.

\textbf{In the online phase}, \Octopus{} uses load-balanced scheduling to retrieve the selected clusters efficiently under the multi-SSD layout with negligible overhead. It also dynamically adapts cluster membership and caching decisions as decoding progresses, thereby sustaining high bandwidth utilization and retrieval efficiency under evolving access patterns.

We evaluate \Octopus{} on four real-world datasets with two types of SSDs, benchmarking five representative LLMs. The results demonstrate that \Octopus{} achieves an average $2.41\times$ reduction in I/O time and an average $2.72\times$ improvement in effective bandwidth utilization. Notably, when scaled to eight SSDs, \Octopus{} attains a bandwidth of up to $37.67$ GB/s, comparable to that between GPU HBM and CPU DRAM.

We summarize the contributions of this paper as follows:
\begin{itemize}
    \item We identify a prevalent yet underexplored phenomenon in LLM inference, KVCache Co-Activation, and show how it can be exploited to mitigate the primary bandwidth bottleneck in SSD-based KVCache offloading.
    \item We design a two-phase solution that systematically models, places, retrieves, and updates co-activated KVs across SSDs, maximizing bandwidth utilization.
    \item We conduct extensive evaluations across diverse LLMs, datasets and hardware configurations, achieving performance and cost improvements over existing solutions.
\end{itemize}

\section{Background}
\subsection{KVCache in Autoregressive Generation}
State-of-the-art LLMs predominantly adopt autoregressive decoding, where tokens are generated sequentially and each new token attends to all previous tokens. This mechanism is implemented by \textit{multi-head attention} (MHA)~\cite{attention}. For each attention head, the input token embedding is projected into three components: Query ($Q$), Key ($K$), and Value ($V$). The attention computation is defined as follows:
\begin{equation}
    \text{Attention}(Q, K, V) = \text{softmax}(\frac{QK^{T}}{\sqrt{d_{k}}})V
\end{equation}
where $d_{k}$ denotes the dimensionality of the key vectors.

At decoding step $t$, the current query $q_t$ attends to all previously generated keys ($K_{<t}$) and values ($V_{<t}$). Since these historical tensors remain unchanged once computed, recomputing them at every step results in considerable and unnecessary overhead. This redundancy severely inflates response-generation latency, complicating adherence to service-level objectives (SLOs)~\cite{mooncake}. To eliminate this inefficiency, modern inference systems store and reuse previously computed keys and values, collectively referred to as the \textit{KVCache}.

\begin{table}[b]
    \centering
    \setlength{\tabcolsep}{3pt}
    \caption{KVCache memory footprint relative to the model weights of GPT-3 (in FP16) across sequence lengths.}
    \label{tab:kvcache-memory}
    \small
    \begin{tabular}{c|lllll}
        \toprule
        Weights & $s=32K$      & $s=64K$      & $s=128K$     & $s=256K$      & $s=512K$ \\ \midrule
        326 GB  & 144 GB       & 288 GB       & 576 GB       & 1.13 TB       & 2.25 TB  \\
        -       & 0.44$\times$ & 0.88$\times$ & 1.77$\times$ & 3.53$\times$  & 7.07$\times$ \\
        \bottomrule
    \end{tabular}
\end{table}

In practice, KVCache exhibits two defining features:

\noindent\textbf{Feature \#1: Massive Footprint.} For a model with $L$ layers and hidden dimension $d$, the KVCache memory footprint for a sequence of length $S$ scales as $O(L \times S \times d)$. As models become deeper and context windows expand to accommodate long documents, code repositories, and multimodal inputs (e.g., images and videos), the KVCache can easily exceed the size of the model parameters themselves. As shown in Table~\ref{tab:kvcache-memory}, under long-context settings, the total KVCache footprint may reach hundreds or thousands of gigabytes, imposing severe memory pressure during LLM inference.

\begin{figure}[t]
    \centering
    \includegraphics[width=1.0\linewidth]{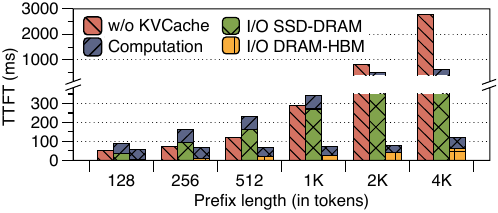}
    \caption{TTFT breakdown for Qwen3-4B with no KVCache, SSD-based KVCache, and DRAM-based KVCache.}
    \label{fig:kvcache-ttft}
\end{figure}

\noindent\textbf{Feature \#2: Temporal Persistence.} The lifetime of the KVCache extends beyond a single decoding step and often even beyond an individual request. On the one hand, a long document or video may be queried repeatedly with different questions while the underlying context remains unchanged. On the other hand, real-world workloads, such as multi-turn agents and iterative coding assistants, continuously reuse historical context across successive interactions. In both cases, discarding and recomputing the KVCache not only wastes computation but also introduces avoidable latency. As shown in Figure~\ref{fig:kvcache-ttft}, effectively reusing the KVCache can significantly reduce time-to-first-token (TTFT), particularly in long-context settings where prefix processing dominates latency.

\subsection{Three-Tier KVCache Offloading}
The massive footprint and temporal persistence of KVCache challenge existing memory hierarchies. In practice, GPU high-bandwidth memory (HBM) serves as the \textbf{first tier}, providing low latency and high throughput, but with limited capacity and high cost. To extend available memory pool, CPU DRAM is commonly employed as a \textbf{secondary tier} for offloading, offering an order of magnitude greater capacity than HBM. During inference, KVCache entries are transferred on demand from CPU to GPU through the PCIe interface. Thanks to the high bandwidth of modern PCIe links~\cite{pcie} and software-level optimizations such as prefetching~\cite{kvcache-prefetch-1,kvcache-prefetch-2} and caching~\cite{adapt-cache,attention-store}, this data transfer overhead is generally manageable.

Nevertheless, DRAM capacity remains fundamentally constrained and expensive at scale. As KVCache grows in both size and lifetime, relying solely on HBM and DRAM becomes economically and practically unsustainable. In this context, solid-state drivers (SSDs) are emerging as a cost-effective \textbf{third tier} for KVCache offloading. Built on NAND flash memory~\cite{flash}, SSDs provide much larger capacity than DRAM, scaling to terabyte levels at lower cost per bit. Moreover, modern server platforms support multiple SSDs to be attached to a single CPU via independent PCIe lanes, enabling flexible capacity expansion and device-level parallelism.

However, incorporating SSDs into the memory hierarchy shifts the bottleneck to \textbf{I/O bandwidth}. Although modern SSDs leverage high-performance protocols such as non-volatile memory express (NVMe)~\cite{nvme}, their bandwidth remains structurally limited compared to DRAM. When KVCache resides in DRAM, it can be transferred to GPU over a PCIe$\times$16 link with relatively high throughput. In contrast, most NVMe SSDs are connected via PCIe$\times$4, offering only a fraction of the bandwidth available between DRAM and HBM. As shown in Figure~\ref{fig:kvcache-ttft}, offloading KVCache to SSDs can introduce substantial I/O latency, and naive designs may severely degrade overall inference performance.

\begin{figure}[t]
    \centering
    \includegraphics[width=1.0\linewidth]{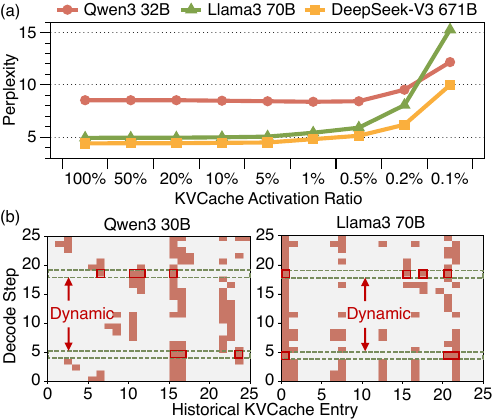}
    \caption{KVCache activation sparsity. (a) Perplexity on WikiText (2K context) under varying KV activation ratios. (b) Dynamic activation patterns of KVCache across decoding steps.}
    \label{fig:kvcahce-sparsity}
\end{figure}

\subsection{Sparsity-Driven KVCache Activation}
To mitigate the overhead of KVCache transfers across memory tiers, recent studies~\cite{h2o,quest,infllm,pqcache} exploit the inherent sparsity within the attention mechanism. Despite large context lengths, only a small subset of tokens with the highest attention scores dominate the final attention output. As shown in Figure~\ref{fig:kvcahce-sparsity}(a), a large proportion of the KVCache can be safely ignored with negligible impact on model outputs. Therefore, instead of activating and transferring all historical keys and values, it is sufficient to identify and load only the KVCache entries of these critical tokens at each decoding step.

However, sparsity-based KVCache activation presents two key challenges. \textbf{First}, the KVCache sparsity pattern is inherently dynamic and query-dependent. As shown in Figure~\ref{fig:kvcahce-sparsity}(b), the set of activated KV entries varies across input tokens. The system needs to compute relevance scores (e.g., query-key inner products) and retrieve critical entries on the fly at each step, requiring a highly efficient retrieval mechanism. \textbf{Second}, although sparsity reduces both data transfers and attention computation, thereby lowering overall inference latency, it does not fundamentally eliminate the SSD bandwidth bottleneck. The improvement arises from shrinking the workload, rather than increasing effective I/O throughput. As a result, SSD bandwidth remains on the critical path of inference and continues to be the dominant performance constraint.

\begin{figure}[t]
    \centering
    \includegraphics[width=1.0\linewidth]{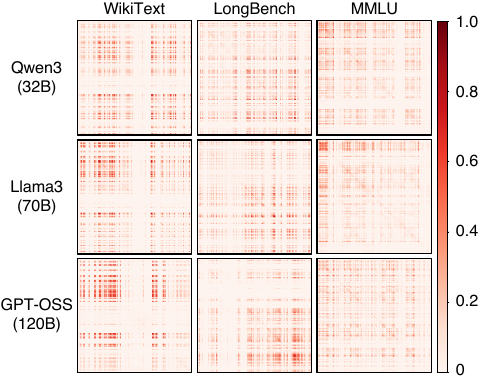}
    \caption{Visualization of KVCache co-activation patterns across models and datasets. Each matrix element $(i, j)$ denotes the activation frequency of KVCache entries $e_i$ and $e_j$.}
    \label{fig:kvcache-co-activation}
\end{figure}

\section{Motivation and Challenges}
\subsection{Insight: KVCache Co-Activation}
\noindent\textbf{Observation.} In this paper, we aim to address the SSD bandwidth bottleneck to enable efficient and scalable KVCache offloading. Our approach is grounded in a profound yet largely underexplored observation: KVCache activation is not uniformly random but exhibits strong correlations. As shown in Figure~\ref{fig:kvcache-co-activation}, when processing real-world workloads, the activation of a given KV entry is often accompanied by a stable and recurring set of other entries. We term this phenomenon \textbf{\textit{KVCache Co-Activation}}, which naturally arises from sparsity-driven KVCache activation and consistently appears across different models and datasets. This observation suggests an opportunity to exploit the structured access patterns.

\noindent\textbf{Core Idea.} KVCache co-activation reveals intrinsic parallelism in KVCache activation patterns. Within each KVCache co-activation group, KV entries are computationally independent and can therefore be fetched concurrently without introducing additional synchronization. This parallelism aligns naturally with the topology of modern storage systems, where multiple SSDs are connected via independent PCIe lanes. By \textbf{\textit{strategically distributing co-activated KV entries across multiple SSDs}}, the system can aggregate device-level bandwidth for KVCache transfers. The resulting aggregated bandwidth can match or even exceed the effective bandwidth between DRAM and HBM, fundamentally alleviating the bottleneck.

\begin{figure}[t]
    \centering
    \includegraphics[width=1.0\linewidth]{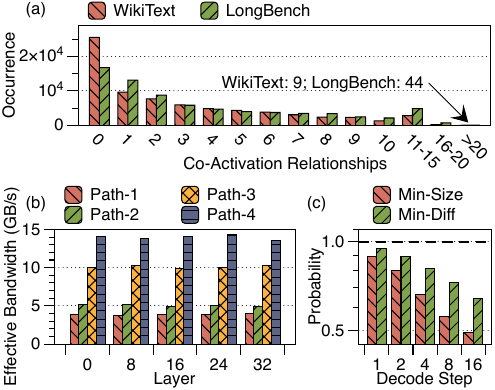}
    \caption{(a) KVCache co-activation distribution on Qwen3-32B over 128 queries. (b) Retrieval paths' impact on effective bandwidth across layers. (c) KVCache co-activation probability degradation across decoding steps under naive updates.}
    \label{fig:challenge}
\end{figure}

\subsection{Main Challenges}
Although KVCache co-activation exposes intrinsic I/O parallelism, translating this insight into a practical SSD-based system is far from straightforward. Based on a comprehensive analysis, we identify four key challenges that must be tackled:

\noindent\textbf{Challenge \#1: Modeling.} The large number of KVCache entries and their complex activation patterns make identifying co-activated entries challenging. As depicted in Figure~\ref{fig:challenge}(a), co-activation patterns in practice are not neatly separable but instead form overlapping structures, with some entries participating in more than 20 co-activation relationships. Selective replication can help decouple these dependencies, but it introduces a combinatorial optimization problem that is NP-hard.

\noindent\textbf{Challenge \#2: Placement.} Beyond identification, the system must further organize these correlated KV entries into a well-structured layout. In particular, this placement should align naturally with the SSD offloading hierarchy. On the one hand, co-activated entries should be distributed across SSDs as evenly as possible to maximize parallel read bandwidth. On the other hand, the system should be able to efficiently select these entries without first transferring them to DRAM.

\noindent\textbf{Challenge \#3: Retrieval.} Replication further complicates retrieval. The same KV entry may reside on multiple SSDs, creating multiple feasible retrieval paths. As shown in Figure~\ref{fig:challenge}(b), different retrieval paths can result in markedly different effective bandwidth, even under an identical KVCache placement. A suboptimal policy may reintroduce load imbalance and negate the benefits of replication, while overly complex scheduling may incur prohibitive per-step overhead.

\noindent\textbf{Challenge \#4: Update.} LLM decoding continuously generates new KV entries that have to be integrated into the existing organization. As shown in Figure~\ref{fig:challenge}(c), naive updates (based on group size or embedding similarity) can severely degrade co-activation probability, limiting achievable parallelism over time. Moreover, caching decisions should adapt to evolving access patterns to reduce repeated I/O for hot entries.

\section{Design Overview}
\begin{figure}[t]
    \centering
    \includegraphics[width=1.0\linewidth]{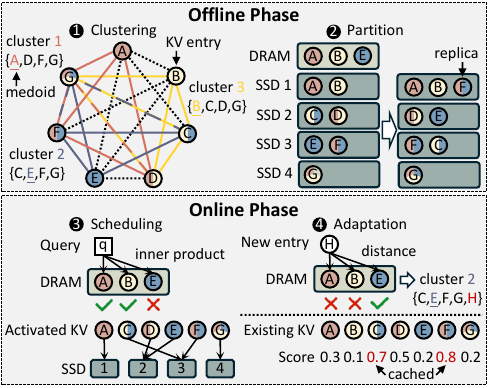}
    \caption{Overview of \Octopus{}.}
    \label{fig:overview}
\end{figure}

Figure~\ref{fig:overview} presents an overview of \Octopus{}. Leveraging KVCache co-activation, \Octopus{} provides a systematic solution for modeling, placing, retrieving, and updating co-activated KV entries. The overall workflow is divided into two phases:

\noindent\textbf{Offline Phase (\S~\ref{sec:offline}).} In this phase, \Octopus{} determines the placement of KV entries across SSDs to maximize aggregate bandwidth utilization. \ding{182} It first models the co-activation patterns of KVCache and organizes entries into clusters accordingly. \ding{182} Meanwhile, \Octopus{} constructs an index structure over clusters to support efficient retrieval during online inference. Based on this clustering, it distributes the entries of each cluster evenly across SSDs to facilitate parallel data transfer.

\noindent\textbf{Online Phase (\S~\ref{sec:online}).} Given the aggregate-oriented placement, \Octopus{} further introduces runtime strategies to fully utilize the available bandwidth. On the one hand, \ding{184} it employs a load-balanced scheduling mechanism to route tokens during cluster retrieval. On the other hand, \ding{185} \Octopus{} dynamically adapts clustering and caching decisions during decoding to maintain high efficiency under evolving access patterns.

\section{Offline Phase: Modeling and Placement}\label{sec:offline}
\subsection{Correlation-Aware Clustering}
\begin{figure*}
    \centering
    \includegraphics[width=1.0\linewidth]{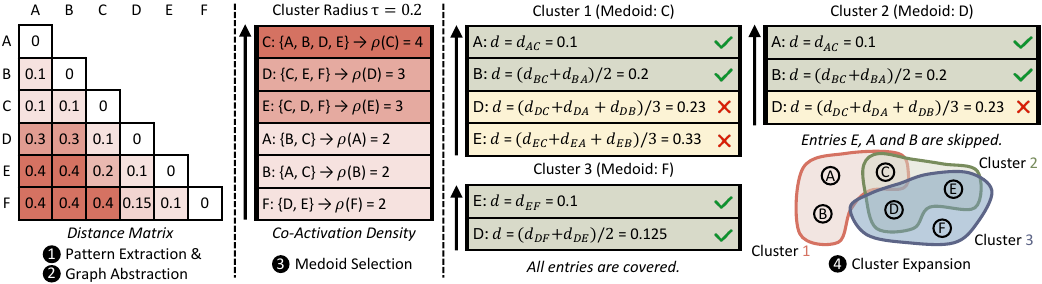}
    \caption{Example of the offline clustering. (1–2) A distance matrix derived from KVCache co-activation patterns forms a graph abstraction. (3) Entries with the highest co-activation density are sequentially selected as medoids. (4) Each cluster expands by iteratively adding entries whose average distance to cluster members is within the radius threshold $\tau$, until all entries are covered.}
    \label{fig:method-offline-clustering}
\end{figure*}

KVCache co-activation analysis reveals that KV entries are typically activated in highly correlated groups rather than independently. Motivated by this insight, \Octopus{} organizes the KVCache into a set of structured \textit{clusters}, each comprising entries with strong co-activation relationships. As shown in Figure~\ref{fig:method-offline-clustering}, \Octopus{} adopts a four-step clustering workflow.

\noindent\textbf{Step 1: Pattern Extraction.} \Octopus{} begins by profiling KVCache co-activation patterns as a one-time preprocessing step prior to inference. For each model layer, we construct an \textit{adjacency matrix} $\mathbf{A}$ to capture pairwise co-activation statistics among KV entries. Each element $\mathbf{A}_{ij}$ records the number of times KV entries $e_i$ and $e_j$ are activated together during attention computation. Thus, $\mathbf{A}$ encodes the empirical co-activation frequency for all entry pairs. Based on the measured frequency $f(e_i,e_j)$, we estimate the co-activation probability between $e_i$ and $e_j$, denoted as $P(e_{i},e_{j})$, as follows:
\begin{equation}
    P(e_{i},e_{j}) = \frac{f(e_{i},e_{j})}{\sum_{k=1}^{N}{\sum_{l=1}^{N}}{f(e_{k},e_{l})}}
\end{equation}
where $N$ denotes the context length. Profiling is performed on a task-agnostic dataset to obtain statistically representative characterizations of KVCache behavior.

\noindent\textbf{Step 2: Graph Abstraction.} Following the extracted co-activation patterns, \Octopus{} abstracts these relationships into a \textit{complete graph} to enable principled graph-theoretic optimization. In this formulation, each node represents a KV entry, and each edge represents the co-activation relationship between a pair of entries. We define the edge weight as a distance metric between two entries, denoted as $d(e_{i}, e_{j})$, which quantifies the strength of their co-activation as follows:
\begin{equation}
    d(e_{i},e_{j}) \coloneq 1 - P(e_{i},e_{j})
\end{equation}
Under this definition, a smaller distance corresponds to stronger co-activation. In the extreme case where $d(e_{i},e_{j}) = 1$, the two entries are always co-activated. Based on this abstraction, a pairwise \textit{distance matrix} $\mathbf{D}$ can be constructed for all entries. Consequently, clustering co-activated KV entries is reformulated as a graph optimization problem: identifying tightly connected node sets with minimal-distance edges.

\Octopus{} completes this optimization procedure in the final two steps. Considering exact combinatorial optimization over a complete weighted graph is NP-hard~\cite{np-hard}, \Octopus{} adopts a heuristic algorithm to balance efficiency and solution quality, as outlined in Algorithm~\ref{algo:clustering}. The core idea is to first select the entry with the highest overall co-activation as the \textit{medoid} (Lines 6-10), and then greedily expand this medoid into a cluster by iteratively incorporating additional entries with strong co-activation relationships (Lines 11-28).

\begin{algorithm}[t]
\small
\caption{KVCache Co-Activation Clustering}
\label{algo:clustering}
\begin{algorithmic}[1]
\State \textbf{Input}: KV entry set $\mathcal{E}$, distance matrix $\mathbf{D}$, cluster radius $\tau$
\State \textbf{Output}: KVCache cluster set $\mathcal{C}$
\Function{BuildClusters}{$\mathcal{E}$, $\mathbf{D}$, $\tau$}
    \State Initialize $\mathcal{C} \leftarrow \emptyset$
    \State Initialize $\text{covered}[e] \leftarrow \mathrm{false}$ for all $e \in \mathcal{E}$
    
    \State $\triangleright$ \textit{Step 3: Medoid Selection}
    \For{each $e_{i} \in \mathcal{E}$}
        \State co-activation density $\rho[e_{i}] \leftarrow \sum_{e_{j} \in \mathcal{E}} \mathbb{I}(\mathbf{D}(e_{i},e_{j}) \leq \tau)$
    \EndFor
    \State medoid queue $Q_{m} \leftarrow$ sort $\mathcal{E}$ in descending order of $\rho[\cdot]$
    
    \State $\triangleright$ \textit{Step 4: Cluster Expansion}
    \For{each $m_{i} \in Q_{m}$ \textbf{and} $\text{covered}[m_{i}] = \mathrm{false}$}
        \State KVCache cluster $C_{i} \leftarrow \{m_{i}\}$
        \State candidate queue $Q_{c} \leftarrow \{e \in \mathcal{E} \setminus \{m_{i}\} \mid D(m_{i},e) \le \tau\}$
        \State $Q_{c} \leftarrow$ sort $Q_{c}$ in ascending order of $\mathbf{D}(m_{i},\cdot)$
        \For{each $c_{j} \in Q_{c}$ \textbf{and} $\sum_{e_{k} \in C_{i}} \mathbf{D}(c_{j}, e_{k})/|C_{i}| \le \tau$}
            \State $C_{i} \leftarrow C_{i} \cup \{c_{j}\}$
        \EndFor
        \State $\mathcal{C} \leftarrow \mathcal{C} \cup \{C_{i}\}$
        \State covered[$e_{i}$] $\leftarrow \mathrm{true}$ for all $e_{i} \in C_{i}$
        \State \textbf{break} if $\forall e \in \mathcal{E},\ \text{covered}[e] = \mathrm{true}$
    \EndFor
    \State \textbf{return} $\mathcal{C}$
\EndFunction
\end{algorithmic}
\end{algorithm}

\noindent\textbf{Step 3: Medoid Selection.} To quantify the co-activation strength of a KV entry, we further define the \textit{co-activation density} of entry $e_{i}$, denoted as $\rho(e_{i})$, as follows:
\begin{equation}
    \rho(e_{i}) \coloneq \bigl|\{e_{j} \mid j\ne i,\; d(e_{i},e_{j})<\tau\}\bigr|
\end{equation}
where $\tau$ is a predefined \textit{cluster radius} that controls the expected cluster size. Under this formulation, an entry with more neighbors within radius $\tau$ has a higher co-activation density, indicating stronger co-activation relationships. \Octopus{} then computes $\rho(e_{i})$ for all KV entries and sorts them in descending order to construct a medoid queue $Q_{m}$. Entries are then selected sequentially from $Q_{m}$ as cluster medoids. By construction, these medoids exhibit the highest co-activation density and thus have the greatest likelihood of co-activating other entries within their respective clusters when activated.

\noindent\textbf{Step 4: Cluster Expansion.} When a medoid $m_{i}$ is selected, \Octopus{} computes its distance to all other entries. Any entry $e_{j}$ satisfying $d(m_{i},e_{j})$ is regarded as a candidate member of cluster $C_{i}$. These candidates are sorted according to their distances to the medoid and organized into a candidate queue $Q_{c}$. For each candidate $c_{j}$ fetched from $Q_{c}$, we define its distance to cluster $C_{i}$ as the average pairwise distance:
\begin{equation}
    d(c_{j},C_{i}) \coloneq \frac{\sum_{e_{k}\in C_{i}}{d(c_{j},e_{k})}}{|C_{i}|}
\end{equation}
If this distance is smaller than $\tau$, $c_{j}$ is added to $C_i$. The cluster is iteratively expanded under this criterion until no further entries can be incorporated. The clustering process terminates once every entry belongs to at least one cluster.

Importantly, the algorithm naturally creates replicas of highly co-activated KV entries, which is crucial for capturing interleaved co-activation patterns. Consider the case where entry $A$ is frequently co-activated with both entry $B$ and entry $C$, while $B$ and $C$ themselves are rarely co-activated. If each entry is restricted to a single assignment, $A$ must either be clustered together with both $B$ and $C$, or grouped with only one of them. In the former case, retrieving the cluster would incur redundant transfer of entry $C$ when only $B$ is needed. In the latter case, the cluster that does not include $A$ would lose the opportunity to benefit from its frequent co-activation with $A$. By allowing replicas of highly co-activated entries, \Octopus{} resolves this tension with negligible storage overhead.

\subsection{Offloading-Friendly Partition}\label{sec:offline-partition}
\begin{figure}[t]
    \centering
    \includegraphics[width=1.0\linewidth]{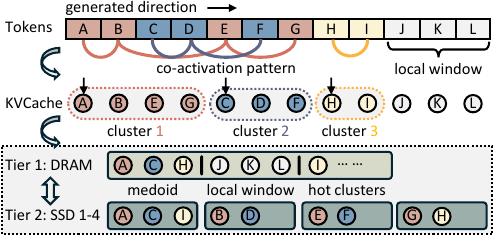}
    \caption{KVCache partition across the DRAM-SSD hierarchy. Medoids, local-window entries, and hot clusters remain in DRAM, while clustered entries are distributed across SSDs.}
    \label{fig:method-offline-placement}
\end{figure}

\Octopus{} treats KVCache clusters as the atomic retrieval units and transfers them from SSD to DRAM on demand. To enable accurate and efficient retrieval, \Octopus{} builds an index structure over KVCache clusters, drawing on the conceptual parallel between KVCache retrieval and classical data retrieval in database systems. By distributing the entries of each cluster across multiple SSDs, the transfer of a single cluster can leverage the aggregated bandwidth of multiple devices, thereby mitigating the bandwidth bottleneck. As depicted in Figure~\ref{fig:method-offline-placement}, \Octopus{} realizes this design through careful partitioning across the two-tier memory hierarchy, minimizing I/O overhead while maximizing aggregate bandwidth utilization.

\noindent\textbf{Tier 1: DRAM.} \Octopus{} places three types of data in DRAM. (1) \textit{Medoids}. \Octopus{} uses the medoid of each cluster as its index entry and maintains a route table that records its SSD placement. Each medoid computes a relevance score with the incoming query to determine whether the corresponding cluster should be retrieved. During clustering, each entry in a cluster is constrained to lie within the radius of its medoid. Consequently, the medoid naturally serves as the representative entry of the cluster and is most likely to be co-activated with other entries in the same cluster. By retaining this index in DRAM, \Octopus{} enables efficient cluster selection without incurring SSD access overhead. (2) \textit{Local window entries}. \Octopus{} keeps KV entries within a local window (e.g., 256 tokens) in DRAM to preserve contextual continuity during inference. A sliding buffer dynamically tracks the most recent entries as generation progresses. (3) \textit{Hot clusters}. To exploit cluster-level activation locality, \Octopus{} caches the most frequently activated clusters in DRAM. To balance space overhead against I/O reduction, we define a \textit{cost-effectiveness score} for cluster $C_{i}$ as follows:
\begin{equation}
\label{eq:cache-score}
    S(C_{i}) = f_{i}\cdot\frac{T_{\text{base}}+s_{i}\cdot T_{\text{transfer}}}{s_{i}}
\end{equation}
where $f_i$ denotes the activation frequency of cluster $C_i$, $T_{\text{base}}$ is the SSD addressing latency, and $T_{\text{transfer}}$ is the per-entry transfer cost. Given a fixed DRAM buffer capacity, clusters are cached in descending order of this score to maximize overall cost-effectiveness under memory constraints.

\noindent\textbf{Tier 2: SSD.} \Octopus{} stores all KVCache clusters across SSDs. Since KV entries have already been clustered based on co-activation patterns, the primary objective here is to distribute clusters as evenly as possible to maximize aggregate bandwidth utilization. \Octopus{} adopts a round-robin placement strategy that flexibly adapts to varying cluster sizes and SSD numbers. Specifically, \Octopus{} maintains a global disk pointer $p_{\text{global}}$ and processes clusters sequentially. For each cluster $C_{i}$, the target disk identifier is determined as:
\begin{equation}
    \text{id}_i = p_{\text{global}} \bmod N_{\text{disk}}, \
    p_{\text{global}} = p_{\text{global}} + |C_{i}|
\end{equation}
where $N_{\text{disk}}$ denotes the total number of SSDs. Each cluster then places its entries starting from disk $\text{id}_{i}$, distributing them sequentially across SSDs in a wrap-around manner. This strategy ensures balanced storage utilization and maximizes parallel data transfers during cluster retrieval.

\section{Online Phase: Retrieval and Update}\label{sec:online}
\subsection{Load-Balanced Scheduling}
After the offline optimization in \Octopus{}, KV entries are placed using a bandwidth-aggregated layout across SSDs. At runtime, \Octopus{} focuses on efficiently transferring the selected clusters based on the incoming query. The design is guided by two key objectives. First, the clustering process naturally introduces KV entry replicas. Although replication is essential for accurately capturing co-activation patterns, it may lead to redundant data transfers if multiple clusters containing overlapping entries are retrieved simultaneously. Second, selected clusters should be assembled in a load-balanced manner to fully utilize the available aggregate bandwidth. Meanwhile, the overall scheduling mechanism must remain lightweight to avoid introducing additional runtime overhead.

\begin{figure}[t]
    \centering
    \includegraphics[width=1.0\linewidth]{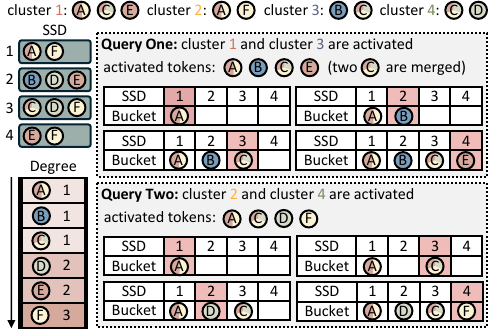}
    \caption{Example of KVCache retrieval scheduling under two queries. Activated clusters are merged, and entries are scheduled across SSD buckets in ascending order of degree.}
    \label{fig:method-online-scheduling}
\end{figure}

As illustrated in Figure~\ref{fig:method-online-scheduling}, \Octopus{} develops a decoupled entry-bucket scheduling algorithm. First, before transferring the KV clusters to DRAM, \Octopus{} performs a global merge over all activated clusters to construct the I/O entry set:
\begin{equation}
    \mathcal{E}_{\text{I/O}} = \left(\bigcup_{C_{i} \in \mathcal{C_{\text{activated}}}}{C_i}\right) \setminus \mathcal{E}_{\text{DRAM}}
\end{equation}
This operation eliminates duplicate KV entries across clusters and filters out entries that are already resident in DRAM, thereby avoiding redundant data transfers. Next, each SSD is assigned a dedicated bucket that stores the entries pending transfer. All entries in $\mathcal{E}_{\text{I/O}}$ are first sorted in ascending order according to their replication factor (i.e., the number of SSDs on which the entry is available). Entries without replicas are directly assigned to their corresponding SSD bucket. For entries with multiple replicas, \Octopus{} routes each entry to the SSD whose bucket currently has the smallest bucket size, balancing the load across devices. When multiple SSDs have the same bucket size, one of them is selected arbitrarily.

After the buckets are constructed for each SSD, the system iteratively extracts the head entry from every bucket and aggregates them into a single submission batch. This batch is then issued to the kernel through a single system call. By consolidating multiple I/O requests into large batched submissions, \Octopus{} minimizes the number of submission operations and reduces user–kernel context switches.

\subsection{Cluster-Aligned Adaptation}
\Octopus{} also dynamically adapts to evolving system states during the online phase. Newly generated KVCache entries may alter the clustering structure, while changing activation patterns shift DRAM caching priorities. To address these dynamics, \Octopus{} employs two dedicated update strategies.

\begin{figure}[t]
    \centering
    \includegraphics[width=1.0\linewidth]{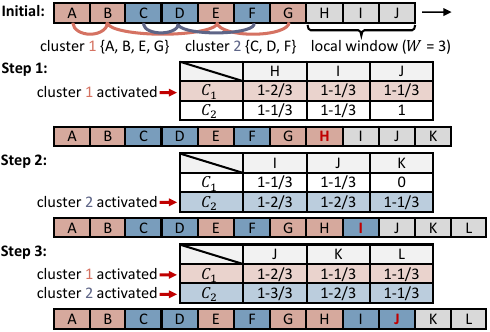}
    \caption{Example of cluster maintenance over three decoding steps. Each new KV entry is assigned to the nearest cluster based on its co-activation pattern in the local window.}
    \label{fig:method-online-adaptation}
\end{figure}

\noindent\textbf{Cluster Maintenance.} As discussed in \S~\ref{sec:offline-partition}, the newly generated KV entries are retained in DRAM within a local window of size $W$. \Octopus{} leverages this $W$-step interval to analyze their co-activation patterns with existing KVCache clusters. As shown in Figure~\ref{fig:method-online-adaptation}, the distance between a newly generated KV entry $e_{new}$ and a cluster $C_{i}$ is defined as:
\begin{equation}
    d(e_{new},C_{i}) = 1 - \frac{f(e_{new},m_{i})}{\sum_{j=1}^{N}{f(e_{new},m_{j})}} = 1 - \frac{f(e_{new},m_{i})}{W}
\end{equation}
where $m_{i}$ denotes the medoid of cluster $C_{i}$ and $N$ is the total number of clusters. If the computed distance is smaller than the cluster radius $\tau$, \Octopus{} assigns $e_{\text{new}}$ to cluster $C_i$ and places it on the next SSD allocated for that cluster (i.e., at disk index $\text{id}_{i} + |C_{i}| + 1$, modulo $N_{\text{disk}}$ if necessary). If multiple clusters satisfy the distance threshold, the entry is added to all of them, preserving controlled replication. This incremental placement strategy maintains cluster coherence while sustaining parallel transfer as decoding progresses.

\noindent\textbf{Cache Replacement.} In the offline phase, \Octopus{} caches KVCache clusters in DRAM according to a cost-effectiveness score. As defined in Equation~\ref{eq:cache-score}, the activation frequency $f_{i}$ used in this score is initially derived from offline profiling. To adapt to evolving runtime behaviors, \Octopus{} dynamically updates the activation frequency of each cluster based on observed access patterns. Specifically, whenever a cluster is activated, its frequency is incremented by one. Conversely, for clusters already cached in DRAM but not activated during a decoding step, their frequencies are decremented by one. This design rewards clusters exhibiting temporal locality while penalizing inactive ones. To efficiently manage the DRAM cache, \Octopus{} maintains the cached clusters using a min-heap structure ordered by their cost-effectiveness scores, enabling efficient updates and replacements.

\section{Implementation}
We implement \Octopus{} as an end-to-end system for efficient SSD-backed KVCache offloading in CPU-GPU collaborative execution. The system spans both Python and C++, comprising more than 2,000 lines of code and covering the full stack from low-level data movement to runtime scheduling.

\noindent\textbf{Asynchronous I/O.} We build a customized C++ I/O backend to fully exploit the aggregate bandwidth of multiple SSDs. For high-performance asynchronous data transfers, we leverage the Linux kernel interface \textit{io\_uring}~\cite{liburing} to issue concurrent, non-blocking read commands. To further eliminate CPU-side copy overhead, all SSD reads are directed into pre-allocated pinned host memory. This allows the NVMe DMA controllers to place data directly into these buffers, enabling zero-copy transfers to GPU HBM via CUDA copy engines~\cite{nvidia-dma}.

\noindent\textbf{Pipelined Prefetching.} To hide SSD access latency during auto-regressive decoding, we design a carefully engineered pipeline that overlaps computation and predictive KVCache transfers. Specifically, while the GPU executes the attention computation for layer $L$, the CPU concurrently performs three stages for layer $L+1$: (1) predicting the KV clusters likely to be accessed by comparing the layer $L$ embedding with the cluster medoids of layer $L+1$; (2) issuing asynchronous fetch requests to load the corresponding data from the SSD into pinned host memory; and (3) streaming the fetched data to GPU HBM through a dedicated CUDA stream. Based on the embedding similarity across adjacent layers~\cite{embedding-similarity}, this design effectively overlaps SSD access with GPU computation.

\section{Evaluation}
\subsection{Experimental Setup} 
\noindent\textbf{Hardware.} Our system comprises an Intel Xeon Gold 6530 CPU (64 cores, 128 threads), 8 NVIDIA H20 GPUs (96 GB HBM each), 1 TB DDR5 DRAM, and up to 8 NVMe SSDs. To cover heterogeneous storage, we use both high-end Samsung PM9A3 SSDs (up to 1.1M IOPS, 6.9 GB/s) and low-end Intel Optane 900P SSDs (up to 0.55M IOPS, 2.5 GB/s).

\noindent\textbf{Models.}  We choose six widely adopted LLMs, as listed in Table~\ref{tab:exp-model}. These models vary in architecture, scale, and KVCache characteristics. All models employ group-query attention (GQA)~\cite{gqa}, with varying group sizes. We evaluate all models in BF16 precision, following common practice.

\begin{table}[t] 
    \small
    \centering
    \caption{Model configurations.}
    \label{tab:exp-model}
    \begin{adjustbox}{width=1.0\linewidth,center}
    \setlength{\tabcolsep}{2pt}
    \begin{tabular}{lllll}
        \toprule
        \textbf{Model}           & \textbf{Size} & \textbf{Attention} & \textbf{Hidden Dim.} & \textbf{Group Size} \\ \midrule
        Qwen3-S~\cite{qwen3}     & 14B           & GQA                & 5120               & 5                   \\         
        Qwen3-M~\cite{qwen3}     & 32B           & GQA                & 5120                & 8                   \\         
        Llama3.1~\cite{llama3-1} & 70B           & GQA                & 8192                & 8                   \\         
        GPT-OSS~\cite{gpt-oss}   & 120B          & GQA                & 2880                & 8                   \\           
        Qwen3-L-MoE~\cite{qwen3} & 235B          & GQA                & 4096                & 16                   \\         
        \bottomrule
    \end{tabular}
    \end{adjustbox}
\end{table}

\noindent\textbf{Datasets.} We evaluate \Octopus{} on a diverse suite of datasets covering plain text, instruction-following and mathematical tasks. These datasets capture a wide range of linguistic structures, enabling a comprehensive evaluation across varying co-activation patterns. To support longer prefix lengths, we concatenate multiple samples when individual sequences are insufficient, constructing extended inputs of up to 1M tokens.

\begin{table}[t]
    \small
    \centering   
    \caption{Dataset configurations.}
    \label{tab:exp-configuration-dataset}
    \begin{adjustbox}{width=1.0\linewidth,center}
    \setlength{\tabcolsep}{2pt}
    \begin{tabular}{ll}
        \toprule
        \textbf{Dataset}               & \textbf{Description}                      \\ \midrule
        WikiText~\cite{wikitext}       & Wikipedia-based language modeling dataset \\
        LongBench~\cite{longbench}     & Long context understanding capabilities   \\
        MMLU~\cite{mmlu}               & Multitask knowledge evaluation benchmark  \\
        GSM8K~\cite{gsm8k}             & Grade-school math problem dataset         \\
        \bottomrule
    \end{tabular}
    \end{adjustbox}
\end{table}

\noindent\textbf{Baselines.} We compare \Octopus{} against three representative KVCache management methods. All methods are evaluated with a sparsity ratio of 10\%. (1) \textit{No Cluster}: KV entries are stored on SSDs without any clustering or indexing structures. To identify the activated KV entries, the entire KVCache is first loaded into DRAM (in chunks when necessary) to compute attention scores, after which the required entries are retrieved from SSDs. For fairness, we report only the time spent on attention score computation and the transfer of the required entries. (2)\textit{InfLLM}~\cite{infllm}: KV entries are organized into fixed-size blocks. Token importance is determined at the block level, and only the selected blocks are retrieved during inference. (3)\textit{PQCache}~\cite{pqcache}: A clustering-based method that groups KV entries using K-means~\cite{k-means}, where cluster centroids are used to identify relevant tokens for retrieval.

\begin{figure*}[t]
    \centering
    \includegraphics[width=1.0\linewidth]{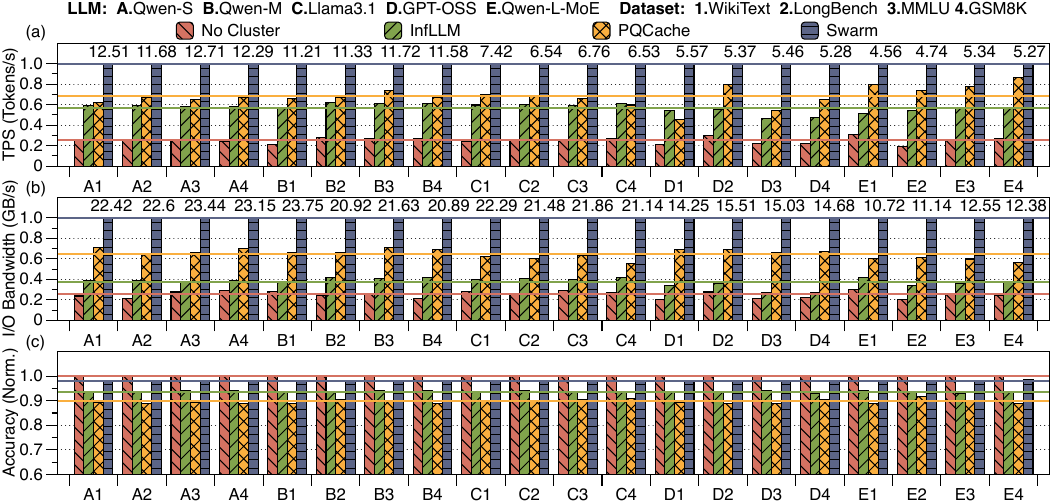}
    \caption{Overall performance (normalized) across models and datasets. Lines indicate average values.}
    \label{fig:exp-overall-performance}
\end{figure*}

\subsection{Overall Performance}
\noindent\textbf{Throughput.} Figure~\ref{fig:exp-overall-performance}(a) reports the end-to-end throughput measured in tokens per second (TPS). \Octopus{} consistently achieves the highest TPS, improving throughput by $3.99\times$, $1.76\times$, and $1.47\times$ over No Cluster, InfLLM, and PQCache, respectively. No Cluster delivers the lowest throughput due to the absence of any structured KVCache management. While InfLLM and PQCache introduce structured approaches, their strategies, based on block partitioning or embedding similarity, do not align well with KVCache co-activation patterns. In contrast, \Octopus{} explicitly clusters KVCache entries according to their activation patterns and distributes them across SSDs, effectively mitigating the bandwidth bottleneck.

\noindent\textbf{Bandwidth.} Figure~\ref{fig:exp-overall-performance}(b) presents the detailed I/O bandwidth during KV retrieval. Consistent with the throughput results, \Octopus{} achieves the highest bandwidth utilization, exceeding No Cluster, InfLLM, and PQCache by $3.95\times$, $2.67\times$, and $1.55\times$, respectively, on average. By combining hierarchical offline and online optimizations, \Octopus{} not only balances the I/O workload across devices but also reduces unnecessary data transfers, approaching the practical bandwidth limits of SSDs. Furthermore, the strong translation of I/O improvements into end-to-end latency reductions confirms that I/O is the primary bottleneck in SSD-backed offloading. \Octopus{} addresses this bottleneck with minimal runtime overhead.

\noindent\textbf{Accuracy.} Figure~\ref{fig:exp-overall-performance}(c) shows the normalized accuracy, where oracle token selection (No Cluster) achieves the highest score. \Octopus{} consistently attains accuracy closest to this upper bound across all models and datasets. This result suggests that \Octopus{} effectively captures KVCache co-activation patterns that closely align with the attention dynamics, enabling it to retrieve the most informative tokens. In contrast, other baselines exhibit a mismatch between KVCache structure and activation patterns, leading to the inclusion of irrelevant tokens or the exclusion of important ones.

\subsection{Ablation Study}
\begin{figure}[t]
    \centering
    \includegraphics[width=1.0\linewidth]{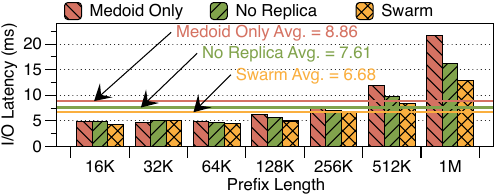}
    \caption{I/O latency comparison across different prefix lengths for Qwen3-32B under three clustering strategies.}
    \label{fig:exp-ablation-offline-cluster}
\end{figure}

\noindent\textbf{Offline Modeling.} We first evaluate the effectiveness of our clustering algorithm. We consider two baselines: \textit{Medoid Only}, which constructs clusters using only entries within the cluster radius of the medoid, and \textit{No Replica}, which restricts one entry to belong to a single cluster. As shown in Figure~\ref{fig:exp-ablation-offline-cluster}, \Octopus{} reduces latency by on average $1.14\times$ and $1.33\times$ compared to these baselines, respectively. These results highlight the importance of capturing pairwise co-activation relationships within clusters and demonstrate that entry replication is essential for modeling complex activation patterns.

\begin{figure}[t] 
    \centering
    \includegraphics[width=1.0\linewidth]{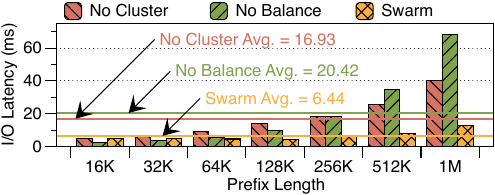}
    \caption{I/O latency comparison across different prefix lengths for Qwen3-32B under three SSD placement strategies.}
    \label{fig:exp-ablation-offline-placement-ssd}
\end{figure}

\noindent\textbf{Offline Placement-SSD.} We compare the aggregate bandwidth achieved by \Octopus{} against two baselines in Figure~\ref{fig:exp-ablation-offline-placement-ssd}. The first, \textit{No Cluster}, places tokens sequentially across SSDs without considering co-activation relationships. The second, \textit{No Balance}, organizes tokens by clusters but places them starting from a single SSD, leading to imbalanced utilization. These results show that \Octopus{} reduces I/O latency by up to $2.63\times$ and $3.17\times$, respectively, effectively translating clustering into higher aggregate bandwidth across SSDs.

\noindent\textbf{Offline Placement-DRAM.} Table~\ref{tab:exp-ablation-offline-placement-dram} evaluates the DRAM placement strategy. In addition to the local window, \Octopus{} keeps cluster medoids resident in DRAM. The results show that \Octopus{} reduces selection latency by $6.9\times$-$12.2\times$ across prefix lengths, with gains increasing proportionally with context length. Importantly, this improvement incurs only minimal memory overhead (0.412 GB at 1M context).

\begin{table}[b]
    \centering
    \setlength{\tabcolsep}{3pt}
    \caption{Selection latency across different prefix lengths on Qwen3-32B under two DRAM placement strategies.}
    \label{tab:exp-ablation-offline-placement-dram}
    \small
    \begin{tabular}{lllll}
        \toprule
        Prefix Len.          & $s=128K$ & $s=256K$ & $s=512K$ & $s=1M$ \\ \midrule
        Naive Lat. (ms)      & 1.635    & 3.136    & 6.176    & 12.15  \\
        \Octopus{} Lat. (ms) & 0.236    & 0.374    & 0.554    & 0.999  \\
        \Octopus{} Mem. (GB) & 0.053    & 0.104    & 0.207    & 0.412  \\
        \bottomrule
    \end{tabular}
\end{table}

\begin{figure}[t] 
    \centering
    \includegraphics[width=1.0\linewidth]{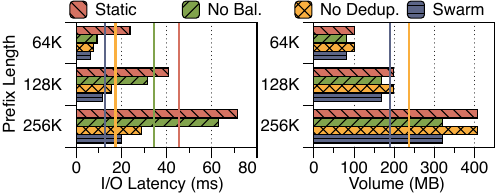}
    \caption{Average I/O latency and volume across different prefix lengths for Qwen3-32B under four retrieval strategies.}
    \label{fig:exp-ablation-online-sched}
\end{figure}

\noindent\textbf{Online Retrieval.} We evaluate the online scheduling algorithm of \Octopus{} in terms of I/O latency and volume under three strategies. \textit{Static} uses static multi-SSD routing, always reading from the first available replica, resulting in neither deduplication nor load balancing. \textit{No Balance} performs deduplication but ignores load balancing, causing requests to concentrate on a subset of SSDs and making the slowest SSD the bottleneck. \textit{No Dedeplication} balances requests across SSDs but does not remove duplicated tokens introduced by cluster replicas, leading to read amplification. Figure~\ref{fig:exp-ablation-online-sched} shows that, by jointly considering replication and load balancing at runtime, \Octopus{} reduces I/O latency by $2.57 \times$ and I/O volume by $1.24\times$ on average compared to all three baselines.

\noindent\textbf{Online Update-Cluster.} Table~\ref{tab:exp-ablation-online-maintenance} evaluates how \Octopus{} maintains cluster quality during decoding. We measure the average distance between cluster entries and their medoid, normalized by the initial distance calculated offline. As baselines, \textit{Min-Size} assigns new tokens based solely on cluster size, while \textit{Min-Diff} assigns them based on similarity to the cluster medoid. \Octopus{} outperforms both baselines by $8.0\times$ and $2.3\times$. By tracking co-activation patterns within a local window, \Octopus{} effectively approximates the evolving relationships between new tokens and existing clusters.

\begin{table}[b]
    \centering
    \setlength{\tabcolsep}{4pt}
    \caption{Distance between cluster entries and medoid (compared to the initial), evaluated on Qwen3-32B with WikiText.}
    \label{tab:exp-ablation-online-maintenance}
    \small
    \begin{tabular}{lllllll}
        \toprule
        Decoding Step & $t=1$ & $t=2$ & $t=4$ & $t=8$ & $t=16$  & $t=32$ \\ \midrule
        Min-Size      & 1.733 & 2.481 & 3.957 & 6.834 & 12.004  & 21.376     \\
        Min-Diff      & 1.581 & 1.880 & 2.184 & 2.348 & 2.649   & 3.019      \\
        \Octopus{}    & 1.001 & 1.001 & 1.001 & 1.002 & 1.005   & 1.012      \\
        \bottomrule
    \end{tabular}
\end{table}

\begin{figure}[t]
    \centering
    \includegraphics[width=1.0\linewidth]{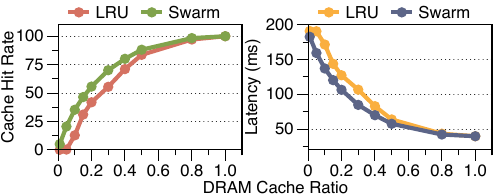}
    \caption{Cache hit rate and end-to-end latency across different cache ratios for Qwen3-32B under Octopus and LRU.}
    \label{fig:exp-ablation-online-cache}
\end{figure}

\noindent\textbf{Online Update-Cache.} Figure~\ref{fig:exp-ablation-online-cache} shows the cache hit rate and per-layer latency under varying DRAM cache ratios. Across all cache budgets, \Octopus{} consistently outperforms the Least Recently Used (LRU)~\cite{lru} baseline, improving cache hit rate by 74\% and reducing per-layer latency by 13.6\% on average. LRU only tracks recent accesses and may keep large clusters that are accessed once but rarely reused. In contrast, \Octopus{} dynamically balances activation frequency and cluster size through a cost-effectiveness score.

\begin{figure}[t]
    \centering
    \includegraphics[width=1.0\linewidth]{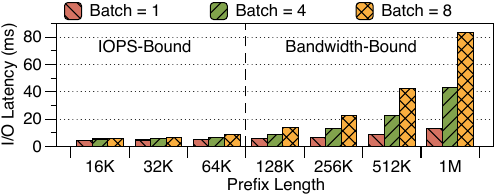}
    \caption{Average I/O latency across prefix length under varying batch sizes for Qwen3-32B on WikiText.}
    \label{fig:exp-sensitivity-length}
\end{figure}

\subsection{Sensitivity Analysis}
\noindent\textbf{Prefix Length.} Figure~\ref{fig:exp-sensitivity-length} presents the I/O latency under different prefix lengths with batch sizes of 1, 4, and 8. For small prefix lengths, latency increases slowly as the workload is IOPS-bound: the number of I/O requests is limited and SSD bandwidth is underutilized. As the prefix length grows, the system becomes bandwidth-bound, with latency scaling approximately linearly with the data retrieved. Increasing the batch size raises the number of concurrent I/O requests, enabling \Octopus{} to enter the bandwidth-bound regime earlier.

\begin{figure}[t]
    \centering
    \includegraphics[width=1.0\linewidth]{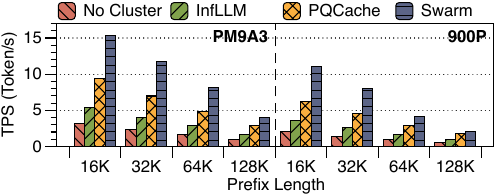}
    \caption{Throughput comparison across different prefix lengths for Qwen3-32B under two SSD types.}
    \label{fig:exp-sensitivity-ssd-type}
\end{figure}

\noindent\textbf{SSD Type.} We evaluate \Octopus{} and the baselines on two storage configurations: four high-tier PM9A3 SSDs and four lower-tier 900P SSDs. Figure~\ref{fig:exp-sensitivity-ssd-type} shows that across all prefix lengths, \Octopus{} consistently achieves higher throughput than all baselines. On the lower-performance SSDs, the system transitions from IOPS-bound to bandwidth-bound at shorter prefixes (around 32K) due to limited IOPS capability. In contrast, with the higher-performance PM9A3 SSDs, this transition occurs at larger prefixes (around 64K), where \Octopus{} achieves a higher upper bound on available bandwidth.

\begin{figure}[t]
    \centering
    \includegraphics[width=1.0\linewidth]{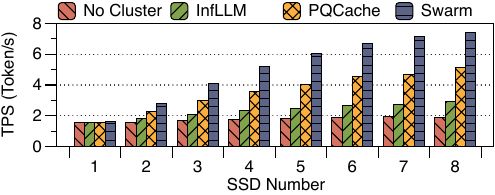}
    \caption{Throughput comparison across different number of SSDs for Qwen3-32B on WikiText (128K prefix length).}
    \label{fig:exp-sensitivity-ssd-number}
\end{figure}

\noindent\textbf{SSD Number.} Figure~\ref{fig:exp-sensitivity-ssd-number} shows throughput as the number of SSDs increases from 1 to 8. With a single SSD, \Octopus{} falls back to the baseline. As the number increases, its throughput scales steadily and consistently surpasses all baselines. This scalability stems from two factors: the offline placement adapts flexibly to varying numbers of SSDs, and the online scheduling effectively balances workloads.

\begin{figure}[t]
    \centering
    \includegraphics[width=1.0\linewidth]{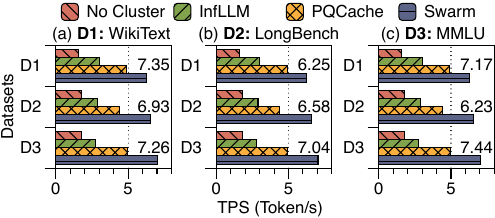}
    \caption{Throughput under dataset-specific $\tau$ calibrated for each datasets, evaluated on Qwen3-32B (128K prefix length).}
    \label{fig:exp-sensitivity-param}
\end{figure}

\noindent\textbf{Clustering Threshold.} We study the sensitivity of throughput to the cluster radius $\tau$ across three datasets, as shown in Figure~\ref{fig:exp-sensitivity-param}. In each subfigure, $\tau$ is selected based on the corresponding dataset. For example, Figure~\ref{fig:exp-sensitivity-param}(a) illustrates the setting where \Octopus{} uses D1 as the offline dataset and is evaluated on D1–D3. The results show that \Octopus{} remains effective even under dataset shift, demonstrating the robustness of our clustering approach.

\begin{figure}[t]
    \centering
    \includegraphics[width=1.0\linewidth]{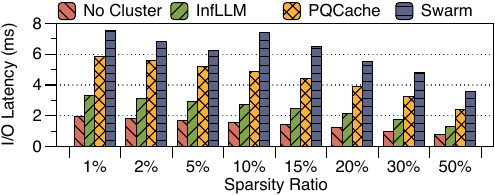}
    \caption{I/O latency comparison across different sparsity ratios for Qwen3-32B on WikiText (128K prefix length).}
    \label{fig:exp-sensitivity-sparsity}
\end{figure}

\noindent\textbf{Sparsity Ratio.} Figure~\ref{fig:exp-sensitivity-sparsity} presents the throughput under varying sparsity ratios. At low sparsity ratios, the I/O volume is small, and performance is primarily limited by IOPS, leading to underutilized bandwidth. As the sparsity ratio increases, the workload becomes bandwidth-bound, with throughput constrained by the maximum achievable I/O bandwidth. Across all settings, \Octopus{} consistently outperforms the baselines. By effectively leveraging aggregated multi-disk bandwidth, \Octopus{} achieves higher throughput in both regimes.

\section{Related Works}
\noindent\textbf{KVCache Sparsity.} Recent work has increasingly explored the intrinsic sparsity of KVCache in long-context LLM inference. Quest~\cite{quest} proposes query-aware sparsity by dynamically selecting relevant KV pages based on the query. RetrievalAttention~\cite{retriAttn} leverages approximate nearest neighbor search to retrieve KV entries via vector similarity. HeadKV~\cite{headkv} introduces head-level sparsity by retaining only a subset of important attention heads. However, a critical remaining challenge is translating these insights into tangible performance gains in real-world inference systems. \Octopus{} bridges this gap with a co-design of algorithms and systems.

\noindent\textbf{KVCache Offloading.} In long-context LLM inference, the KVCache imposes substantial memory pressure, motivating the offloading of KVCache to DRAM and SSD. FlexGen~\cite{flexgen} proposes a three-tier memory hierarchy spanning HBM, DRAM, and SSD, leveraging linear programming to optimize tensor placement. SolidAttention~\cite{solidattention} adopts an interleaved layout of KV entries on SSD and incorporates speculative prefetching, targeting efficient single-node deployments. IMPRESS~\cite{impress} and CachedAttention~\cite{cachedattention} extend this line of work to cloud settings by organizing KV entries into chunks based on token importance within multi-tier storage hierarchies. Despite these advances, prior work identifies SSD bandwidth as the primary bottleneck. In contrast, \Octopus{} addresses this limitation by exploiting multi-disk parallelism.

\noindent\textbf{KVCache Clustering.}
In KVCache offloading, the organization of KV entries plays a critical role in enabling efficient retrieval. ClusterKV~\cite{clusterkv} applies K-means clustering based on static similarity metrics. InfLLM~\cite{infllm} selects high-score tokens as representatives and retrieves KV entries at a coarse granularity. PQCache~\cite{pqcache} instead formulates cluster retrieval as a vector search problem. These methods primarily rely on mathematical similarity or heuristic importance for clustering. \Octopus{} organizes KV entries based on co-activation patterns, translating them into tangible performance gains.

\section{Conclusion}
We propose \Octopus{}, an SSD-based KVCache offloading framework that leverages KVCache co-activation to enable bandwidth-efficient parallel retrieval across SSDs. By transforming bandwidth-bound KVCache access into scalable parallel I/O, \Octopus{} paves the way for cost-efficient and capacity-scalable storage systems for LLM inference.

\bibliographystyle{plain}
\bibliography{references}

\end{document}